\documentclass[manuscript]{aastex}

\slugcomment{To Appear in the Astrophysical Journal Letters}

\shorttitle{X-ray Emission from Abell 3120}
\shortauthors{Mark Henriksen}

\begin{document}

\title{The X-ray View of Abell 3120}

\author{Mark Henriksen \& Alexis Finoguenov}
\affil{Physics Department, University of Maryland, Baltimore County,
    Baltimore, MD 21250}

\begin{abstract}
Identification of Abell 3120 as a galaxy cluster has recently been questioned with 
alternative suggestions including: a fossil remnant
of a group merger, non-thermal emission from a radio galaxy, and projected emission from
of a filamentary string of galaxies. We report on our analysis of the Chandra observation and evaluate
these hypotheses based on our results. Abell 3120 shows X-ray emission extending 158 kpc, well beyond the central galaxy. The spatial distribution
of X-rays in the core more closely
follows the radio emission showing a jet-like structure extending to the north that is misaligned
with the stellar light distribution of the central galaxy.
At larger radii the X-ray emission is aligned with the SE-NW running axis of the galaxy distribution in the
cluster core. Modeling the X-ray spectrum excludes purely non-thermal emission.  The emission weighted temperature is 1.93 - 2.19 keV and the 0.3 - 10 keV luminosity
is 1.23$\times$10$^{43}$ ergs s$^{-1}$. Abell 3120 appears to be a poor cluster with Virgo and MKW 4 as peers. 
The best fitting model
consists of a thermal component and a second component that may be either thermal or non-thermal,
with luminosity $\sim$25\% of the total X-ray luminosity. While, a more detailed spatial-spectral search 
failed to detect a central AGN, there is some evidence for an extended hard X-ray component. Cooler gas, 1.28 - 1.80 was detected in
the central 20 kpc. The second thermal component marginally requires a higher redshift, $>$0.12, 
which may be due to a second cluster in the rich surrounding environment consisting of
nearly a thousand catalogued galaxies.  
\end{abstract}


\section{Introduction}

Abell 3120 is located near the center of the Horologium-Reticulum Supercluster. 
An extended X-ray source was found with ROSAT and identified with the 2MASX J03215645-5119357 galaxy. The emission was characterized as the fossil remnant of galaxy group merger (Romer et al. 2000).
Sparse kinematical data shows a high velocity dispersion around the central galaxy and it was suggested, alternatively, that the
X-ray emission may due to a filament or other line of sight structure (Fleenor et al. 2006). Since there is a radio source identified with the central early-type galaxy,
these authors also suggested that the X-ray emission may be wholly non-thermal and identified with the radio galaxy. 
A detailed study of the X-ray emission with high spatial and spectral resolution can play an important role in evaluating these three hypotheses
and we report here on the X-ray characteristics of Abell 3120 using Chandra observations.

\section{Optical Analysis}

Abell 3120 is classified as a poor cluster (Abell, Corwin, \& Olowin, 1989). However, the region is actually quite rich with 813
galaxies catalogued by the APM survey (Maddox et. al., 1990)  within 3 Mpc of the cluster center. The DSS red band image has been smoothed with a 40 kpc Gaussian (at the redshift of Abell 3120) 
to bring out regions of high surface brightness, including bright galaxies and areas of high galaxy density. Figure 1 shows a contour map overlaid on
the smoothed DSS - R band image, 650 kpc on a side. Substructure is visible in the optical image. 
Within 13.2 arcmin, or 1 Mpc (H$_o$ = 73 km sec$^{-1}$Mpc$^{-1}$, $\Omega_{m}$ = 0.27, $\Omega_{v}$ = 0.73 and using the
central radio galaxy's
redshift of 0.0699 corrected to the reference frame defined by the Cosmic Microwave Background Radiation,
the scale is 1.284 kpc arcsec$^{-1}$) in projection from the Abell 3120 center, there are only 7 galaxies
with measured redshift (see Table 1). The mean velocity for these galaxies is 21236 km s$^{-1}$ with a dispersion
of 2824 km s$^{-1}$. 
The galaxy with the highest velocity
is more than 2$\sigma$ from the mean. If it is removed, then the mean is lowered to 20248 km s$^{-1}$ 
and the dispersion is significantly lowered to 1416 km s$^{-1}$. However, this is still a very high dispersion, even for
a very high mass cluster,
if it is in virial equilibrium. Figure 2 shows the optical contours overlaid on the X-ray image. The radio galaxy associated with the central galaxy is
centered on the X-ray peak. A potential problem with this "clipped-mean" is that it gives 
a very high peculiar velocity for the central galaxy, 792 km s$^{-1}$.  This is a significantly higher peculiar velocity than is typical for central galaxies
associated with clusters showing substructure (Oegerle \& Hill, 2001). If we assume that the radio galaxy 
is at rest in the cluster potential and recalculate the mean velocity relative to it, the 
dispersion is even higher, 1,617 km s$^{-1}$. One must conclude that if these galaxies are representative
of the cluster as a whole, it is unlikely that even the bulk flow from merging substructures could
account for such a large velocity dispersion, given the relatively low mass cluster implied by the X-ray temperature (Section 3.3). 
An alternative hypothesis is that there is more than more cluster in the region sampled by the redshifts.

\section{X-ray Observations and Analysis}

Abell 3120 was observed on June 3, 2006 for approximately 27,130 seconds with the ACIS-I detector.  The event 2 file produced from reprocessing III
was used for both the spectral and spatial analysis. Processing includes corrections due to time dependent gain, charge transfer inefficiency (CTI), dead area correction, and bad pixels. The full energy
band was used to extract the spectrum, however, during spectral fitting the energy band was limited to 0.3 - 10 keV. The mkacisrmf tool was
used to create the response matrix file (rmf). A circular region, centered on the visual X-ray peak with radius 250 pixels or 123 arcsec, using the ACIS 
scale of 0.492 arcsec pixel$^{-1}$, was used to extract the source spectrum.
This region includes all of the X-ray emission visible
in the radial profile (Figure 3).
Three point sources within the source region were excluded. Three background regions were chosen in source
free regions of the same chip as the source to extract a background spectrum.
The analyzed
spectrum for the Abell 3120 cluster has an exposure time of 26,784 seconds and a background subtracted count rate of 0.1645 +/- 0.0032 counts per second.
Both simple and complex models were fit to the spectrum including: (1) thermal, (2) powerlaw,
(3) thermal plus powerlaw, (4) two thermal, (5) thermal with two redshifts, and (6) two
thermal with different redshifts.
In addition, a spatial analysis was done consisting of visually correlating emission in several energy bands,
X-ray, K, R, and radio, as well as fitting radial X-ray surface brightness profiles.

\subsection{Spatial Analysis}

We fit the standard profile, S/$\over$S$_{c}$ = (1 + (r/r$_{c}$)$^{2}$)$^{-3\beta + 0.5}$ to the radial surface brightness
distribution in 3 energy bands: soft (0.3 - 2 keV), hard (2 - 10 keV), and total.
The radial profile for the 0.3 - 10 keV energy band is shown in Figure 3. The best fit values are r$_{c}$ = 7.6 kpc
and $\beta$ = 0.441 with 90\% confidence intervals of 6.9 - 8.4 kpc and 0.434 - 0.448, respectively. These are comparable to MKW 4, which is characterized by 
r$_{c}$ = 4.28 - 4.59 kpc and $\beta$ = 0.446 - 0.448. 
The value of r$_{c}$ and $\beta$ differ significantly from rich clusters, which have
larger beta and larger core radii (Vikhlinin, Forman \& Jones 1999), and is typical of groups of galaxies (Mulchaey et al. 2003).
The radial surface brightness profile shows emission out to 250 pix or 123 arc sec (158 kpc), 
well beyond the central galaxy and out into the cluster potential. Within this region there are 16 catalogued galaxies. Thus, while the X-ray gas may be, in large part, trapped by
the galaxy gravitational potential in the central region, it smoothly connects to the intracluster medium. 
Because there is a radio source associated with the cluster, we fit radial profiles in a soft, 0.3-2 keV band, and a hard, 2 - 10 keV band, to search for non-thermal emission in the X-ray. 
The 90\% confidence intervals are r$_{c}$ = 6.1 - 7.6 kpc and  $\beta$ =  0.433 - 0.451 for the soft band and r$_{c}$ = 6.6 - 14.1 kpc and  $\beta$ =  0.374 - 0.439 for the hard band. 
The central count rate is 8.8 times higher in the soft band compared to the hard band. The predominance of soft emission in the center is reflected in the fit parameter
$\beta$, which is nearly identical to the total band.
The beta model is flat in the center and fits both profiles well. It appears that there is no hard, central point source that can be identified with an AGN.  However, the hard band has a larger core radius and a smaller beta making it significantly flatter than the soft emission.  This can be seen clearly in the
radial profiles (Figures 4 and 5).

\subsection{X-ray Spectral Results}

\subsection{Single Component Models}

While a single thermal component is not the best fitting model for the ACIS spectrum, 
the results (Table 2) are useful for comparison to correlations found in surveys of X-ray emission from galaxies, groups, and clusters. 
The X-ray luminosities and temperatures shown in the correlations with velocity dispersion for groups and clusters (Jeltema et al. 2006; Ortiz-Gil et al. 2004; Zimer, Zabludoff, \& Mulchaey 2003)
range from 5$\times$10$^{44}$ - 10$^{46}$ for clusters with a velocity dispersion above 1000 km s$^{-1}$. This is a factor of almost 50 - 1000 above the X-ray
luminosity for Abell 3120, 1.23$\times$10$^{43}$ ergs s$^{-1}$ (Table 3). 
The temperature of Abell 3120, 1.93 - 2.19 keV, is also lower than predicted by the cluster correlations for a high velocity
dispersion (Jeltema et al. 2006; Ortiz-Gil et al. 2004). Clusters with that temperature range in the correlations have velocity dispersions in the range, 350 - 600 km s$^{-1}$. 
Though only weakly constrained by the X-ray data,
the X-ray determined redshift range, z = 0.069 (20,700 km s$^{-1}$) - 0.087 (26,100 km s$^{-1}$), 
while in agreement with the redshift of the radio galaxy, gives a lower bound that excludes the
dynamical model based on the galaxies, as discussed in section 2, with 90\% confidence. Thus we conclude
that the velocity dispersion is unrelated to the Abell 3120 cluster based on cluster correlations.

Comparison to galaxy group properties show that it is somewhat unlikely that the X-ray emission is from a galaxy group
or single galaxy (fossil remnant of a group merger).
The emission weighted temperature, 1.93 - 2.19 keV and luminosity, 1.23$\times$10$^{43}$ ergs s$^{-1}$
are both high compared to groups (Ponman et al. 1996; Jeltema et al. 2006). Only HCG 62 and the NGC 5044 group have a comparable
luminosity to Abell 3120. However, their temperature is significantly lower.  The distribution of temperature for galaxy clusters shows very few 
that are around 2 keV (Mulchaey et al. 2003; Heldson \& Ponman 2000)

The correlation between cluster X-ray luminosity and temperature spanning groups, poor clusters, and rich clusters (Ortiz-Gil et al. 2004; Jeltema et al. 2006), shows that Abell 3120 is typical of low X-ray luminosity
clusters.
The peer class of objects for Abell 3120 appears to be poor clusters with dominant central galaxies. Two
specific examples, both comparable in their X-ray properties, are MKW 4 and Virgo. 
MKW 4, centered on NGC 4073, is a poor cluster which has a temperature of 1.75 - 1.81 keV
in the central 2 arc min (O'sullivan et al. 2003). MKW 4 has a temperature inversion
in the center and reaches an ambient cluster temperature of $\sim$3 keV so that the emission weighted temperature
of the ambient intracluster medium may be somewhat higher. The MOS spectrum provides an emission weighted temperature of 2.5 - 3.82 keV for this cluster.
Abell 3120 has an average abundance, 0.45 - 0.82 Solar, which is higher than is typical for rich clusters, 0.3 Solar, yet also comparable
to MKW 4, 0.76 - 0.87 (Fukazawa et al. 2004).
The spectrum of both MKW 4 and Abell 3120 is better fit by a two component model with the additional component
being either thermal or powerlaw.

The Rossi X-ray Timing Explorer observation of Virgo
gives a temperature of 2.56 +/- 0.03 keV and an abundance of 0.26 +/- 0.02 Solar (Reynolds, et al. 1999). The emission weighted
temperature breaks down into a two phase distribution in the inner 10 arcmin with one component being isothermal around
1 keV and the second increasing from 1.74 keV in the center to 2.5 keV at around 12 arcmin (Matsushita et al. 2002). The abundance of alpha elements shows a decrease from approximately solar in the center to 
approximately 0.3 solar at 12 arcmin.  Because of the large field of view of the RXTE PCA, its values are more weighted toward the cluster emission than the cooler, high abundance gas
associated with M87, which may account for the lower abundance.

\subsection{Multiple Thermal Components}

 The single thermal component model has a reduced chi-square of 1.17. The probability of
exceeding that value is 5\%, under the assumption that it is a good description of the data. Based
on this low probability, we fit more complex models to the data and obtain the following results
via the F-test. The F value is 13.2, 16.3, and 15.5 for adding either a powerlaw, second
thermal, or second redshift, respectively. This means that all additional parameters (see Table 2) have greater than 99.9\%
significance. Allowing the second thermal component to have an independent redshift has an F
parameter of 2.7 and a significance of 90\%.

For the two thermal component model,
temperatures and luminosities are 1.06 - 1.37 keV (8.9$\times$10$^{42}$ ergs s$^{-1}$) and 2.41 - 4.31 keV (3.9$\times$10$^{42}$ ergs s$^{-1}$). 
The low temperature component and luminosity is consistent with the L-T correlation
while the high temperature component is quite underluminous. Alternatively, the high temperature component may be significantly hotter than that predicted by the correlation.
Because the hot component is relatively small, $\sim$30\% of the total luminosity, it may be a clump of non-virialized, shocked gas from a small mass merger.
On the scale of $\sim$100 kpc, the X-ray emission appears asymmetrical and elongated along the NW-SE axis. This
alignment is similar to the optical light distribution (Figure 2), which traces the galaxy distribution
on a much larger scale. The non-spherical distribution is typical of a non-virialized cluster
that has experienced recent merger and accretion events. 

A comparison of the our two component model parameters for Abell 3120 to those obtained for early-type galaxies
indicates that it is unlikely that the emission is due to a single galaxy. The sample of early-type galaxies observed with the Advanced Satellite for Cosmology and Astrophysics (ASCA)
shows that the two component galaxy model consists of a soft one, with
0.3 keV $<$ kT $<$ 1.0 keV and luminosity between 6.6$\times$10$^{39}$ - 2.6$\times$10$^{42}$ and a harder one (Matsumota et al. 1997) .
The hard component can be fit by a $\sim$10 keV thermal or $\alpha$ $\sim$1.8 powerlaw with luminosity
between 3.0$\times$10$^{40}$ - 3.1$\times$10$^{41}$. Comparison with Abell 3120 shows that
for either two component model: two thermal components or thermal plus powerlaw (next section), the hard component in Abell 3120
exceeds the maximum predicted galactic hard component luminosity by an order of magnitude. Thus, a wholly
galactic origin is unlikely. 
The interpretation of the hard component as shocked gas in the two thermal component model 
is also somewhat problematic since the amount of high temperature gas is quite large when compared to simulations of relatively small merger events.

\subsection{Non-thermal Emission}

A single non-thermal component is a poor fit to the data giving a reduced $\chi^{2}$ of 2.5. Thus, one can conclude
that some of the emission must be thermal. A combination of thermal and power-law emission provides a better fit than a single
thermal and it is an equally good fit to the model with two thermal components. In this
model, the temperature is reduced to 1.30 - 1.78 keV and the luminosity of the thermal
component is reduced to 10$^{43}$ ergs s$^{-1}$.

The powerlaw luminosity is 4.2$\times$10$^{42}$ ergs s$^{-1}$ for the radio source. Comparison
with other X-ray detected radio galaxies shows that the non-thermal component is similar
to the nucleus in 3C 98 as opposed to the lobes, which are a factor of 100 weaker (Isobe et al. 2005).
The radio galaxy PKS 1343-601 (Tashiro et al. 1998) has
a factor of 10 lower non-thermal luminosity from its lobes. Other reported lobe
detections include Fornax A with ASCA at 2.8$\times$10$^{41}$ ergs s$^{-1}$
(Kaneda et al. 1995) and with ROSAT at 2.4$\times$10$^{41}$ ergs s$^{-1}$
(Feigelson et al. 1995). From these comparisons, the non-thermal component in Abell 3120
appears to be higher than is typical of lobes and is therefore likely to be associated with core emission from the radio galaxy. While the spectral
index is not well constrained at the low end, the upperlimit with 90\% confidence is 1.56, also consistent with
core emission.

The SUMMS survey (Mauche et al. 2003) at 843 MHz shows a radio source at the position of the Abell 3120 cluster. The source is located at 03 21 56.27+/- 1.6  -51 19 35.1+/-1.8   
with flux  47.6+/-1.6 mJy. Figure 7 shows the radio contours overlaid on the X-ray
and on the K band images. The radio source
is coincident with the X-ray and K-band emission centers. 
The resolution of the radio image is 45x45 arcsec so that
the fitted major and minor axes quoted in the catalog, 57.1 and 48.2 arcsec,
indicate it may be sightly extended. In fact, the radio image appears to extend nearly
2 arcmin in its longest dimension.
The radio contours show a jet-like feature pointing to the N. The X-ray contours in the
core (Figure 8) also show a "jet-like" structure pointing to the N, aligned with the radio source morhpology. Both
are not aligned with the stellar distribution, which points SE-NW.

Two  spatially resolved regions were analyzed to test if there is non-thermal emission: a central region (r$<$15 arc sec or 20 kpc) and an outer annulus, 15 arc sec $<$ r $<$ 100 arc sec or 126 kpc).
The background subtracted spectrum for the inner region contained 1003 counts while the spectrum for the outer region contained 3,449 source counts. Both spectra were 
adequate to distinguish between thermal and a non-thermal emission. The results are shown in Table 4. In both regions, a thermal component model was a significantly better fit than a non-thermal
component, however, the statistical preference was strongest in the inner region. Thus, it is unlikely that non-thermal emission is from the core of the galaxy, or within 20 kpc, as one would
expect from an AGN. This is consistent with the spatial analysis in Section 3.1 that showed the harder emission to have a flatter distribution and not centrally peaked, as would
be expected from a central AGN.

\subsection{Multiple Redshift Components}

We fit two thermal models that involved multiple redshift components: one with a single thermal and
the other with two thermal components. Both provide an improvement compared to the same model with a single redshift.
Both multiple redshift models give one component with a redshift comparable to that of Abell 3120 while the second
is higher, $>$ 0.12.  The model for two thermal components each with its own redshift has an F
parameter of 2.7 and significance of 90\%. Given the large number of galaxies in this region, the higher redshift component
may be from a second cluster along the line of sight.

\section{Summary}

The X-ray image of the Abell 3120 galaxy cluster obtained with Chandra shows emission centered on the central early-type galaxy
but extending well beyond, to 158 kpc. The X-ray emission appears to be better aligned with the radio morphology on small scales,
suggesting that there may be a non-thermal component of X-ray emission. Dividing the radio profile into a hard and soft band shows
that a point source is not visible in either band, however, the hard emission has a flatter distribution. A spectral analysis shows that the emission from within 20 kpc
is thermal and excludes non-thermal emission as the primary component. Beyond the central galaxy, the X-ray emission is aligned with the SE-NW running axis of the 
asymmetrical galaxy distribution. The emission weighted temperature is 1.93 - 2.19 keV and the 0.3 - 10 keV luminosity
is 1.23$\times$10$^{43}$ ergs s$^{-1}$. Modeling the X-ray spectrum excludes purely non-thermal emission. The best fitting model
consists of a thermal component and a second component that may be either thermal or non-thermal,
with luminosity $\sim$25\% of the total X-ray luminosity. If the second component is non-thermal, it can not be confined to the center and
must have a broad, diffuse spatial morphology, similar to the radio source. Since the central 20 kpc and the region beyond have significantly different temperature, 1.28 - 1.80 and 1.99 - 2.58 keV, respectively,
in the spatial-spectro analysis, this may favor a thermal interpretation for the second spectral component . Abell 3120 has an average abundance, 0.45 - 0.82 Solar, which is slightly higher than is
typical of rich clusters. In its X-ray properties,
Abell 3120 appears to be a poor cluster with Virgo and MKW 4 as peers.

\vfill\eject

\begin{deluxetable}{lccc}
\tabletypesize{\scriptsize}
\tablewidth{0pt}
\tablecolumns{4}
\tablecaption{A3120 Galaxies With Velocities\label{tbl-1}}
\tablehead{\colhead{Galaxy} & \colhead{RA} & \colhead{DEC} & \colhead{Velocity}} 
\startdata
2MASX J03215645-5119357   &     03h21m56.4s & -51d19m36s &  21040 \\
2MASX J03214898-5120197   &     03h21m49.0s & -51d20m20s &  19743 \\
2MASX J03214567-5121156   &     03h21m45.7s & -51d21m16s &  22215 \\
2MASX J03223262-5119548   &     03h22m32.6s & -51d19m55s &  18234 \\ 
2MASX J03223357-5127128   &     03h22m33.6s & -51d27m13s &  21396 \\
2MASX J03230265-5113333   &     03h23m02.6s & -51d13m35s &  18865 \\ 
2MASX J03215485-5132207   &     03h21m54.8s & -51d32m22s &  27159 \\ 
\enddata
\end{deluxetable}

\begin{deluxetable}{lcccccccc}
\tabletypesize{\scriptsize}
\tablewidth{0pt}
\tablecolumns{8}
\tablecaption{Spectral Fits \label{tbl-2}}
\tablehead{\colhead{Model} &  \colhead{n$_{H}$ (cm$^{-2}$)} & \colhead{kT$_{1}$ (keV)} & \colhead{Abundance} 
& \colhead{Redshift$_{1}$} & \colhead{kT$_{2}$ (keV)} & \colhead{$\alpha$} & \colhead{Redshift$_{2}$} & \colhead{$\chi^{2}$/dof}}
\startdata
ray  & 0.026 - 0.069 & 1.93 - 2.19 & 0.45 - 0.82 & 0.069 - 0.087 & - & - & - & 234.74/201 \\
pow  & 	-	     &   -          &     -        &     -          &        -             & - & - &493.99/201 \\
ray + pow & 0.053 - 0.111 & 1.30 - 1.78 & 0.19 - 0.46 & 0.052 - 0.082  & -  &  -1.33 - 1.56 &  - &   220.26/200 \\
ray + ray (one z ) & 0.024 - 0.084 & 1.06 - 1.37 & 0.33 - 0.89 & 0.050 - 0.074 & 2.41 - 4.31 & -  & -  & 217.04/200 \\
ray (two z) & 0.045 - 0.10 & 2.05 - 2.36 & 0.99 - 2.41 & 0.57 - 0.61 &   - & - & 0.067 - 0.083 &  219.29/200 \\
ray + ray (two z) & 0.0017 - 0.055 & 2.05 - 2.85 & 1.24 - 2.54 & 0.053- 0.076 &   1.32 - 2.42 & - & 0.140 - 2.20 & 214.11/199 \\
\enddata
\end{deluxetable}

\begin{deluxetable}{lcccccc}
\tabletypesize{\scriptsize}
\tablewidth{0pt}
\tablecolumns{7}
\tablecaption{Spectral Fits - Luminosities\label{tbl-3
}}
\tablehead{\colhead{Model} &\colhead{Normalization$_{1}$} &\colhead{Flux$_{1}$} & \colhead{Luminosity}
 &\colhead{Normalization$_{2}$} &\colhead{Flux$_{1}$} & \colhead{Luminosity}} 
\startdata
ray   &0.00123 - 0.00139 & 1.25e-12 - 1.22e-12 & 1.22e43 - 1.25e43 & -  & - & - \\
ray + pow  &  0.00108 - 0.00185 & 1.02e-12 & 1.02e43 & 4.91e-7 - 1.97e-5 & 4.15e-13  & 4.17e42 \\
ray + ray (one z) &  0.000191 - 0.00126  & 8.84e-13  & 8.87e42 &  0.000407 - 0.00107 & 3.86e-13 & 3.87e42 \\
ray (two z) & 0.000518 - 0.00093 & 8.4e-13 & 8.4e42  &  0.00045 - 0.00080 & 3.7e-13 & 3.7e42 \\
ray + ray (two z) & 0.00050 - 0.00081 &  & &   0.00023 - 0.00049  & &  \\
\enddata
\end{deluxetable}

\begin{deluxetable}{lccccc}
\tabletypesize{\scriptsize}
\tablewidth{0pt}
\tablecolumns{8}
\tablecaption{Spectral Fits \label{tbl-4}}
\tablehead{\colhead{Region} & \colhead{Model} &  \colhead{n$_{H}$ (cm$^{-2}$)} & \colhead{kT(keV)} & \colhead{Abundance} 
& \colhead{$\chi^{2}$/dof}}
\startdata
Inner & thermal          & 0.026 - 0.074 & 1.28 - 1.80 & 0.37 - 0.99 &  68.78/50 \\
Inner & non-thermal  & 	-	          &   -                &     -                &   105.38/51 \\
Outer & thermal          & 0.022 - 0.085 & 1.99 - 2.58 & 0.35 - 0.84 &  220.26/200 \\
Outer & non-thermal &      -                   &         -           &              -      &276.11/201 \\
\enddata
\end{deluxetable}

\plotone{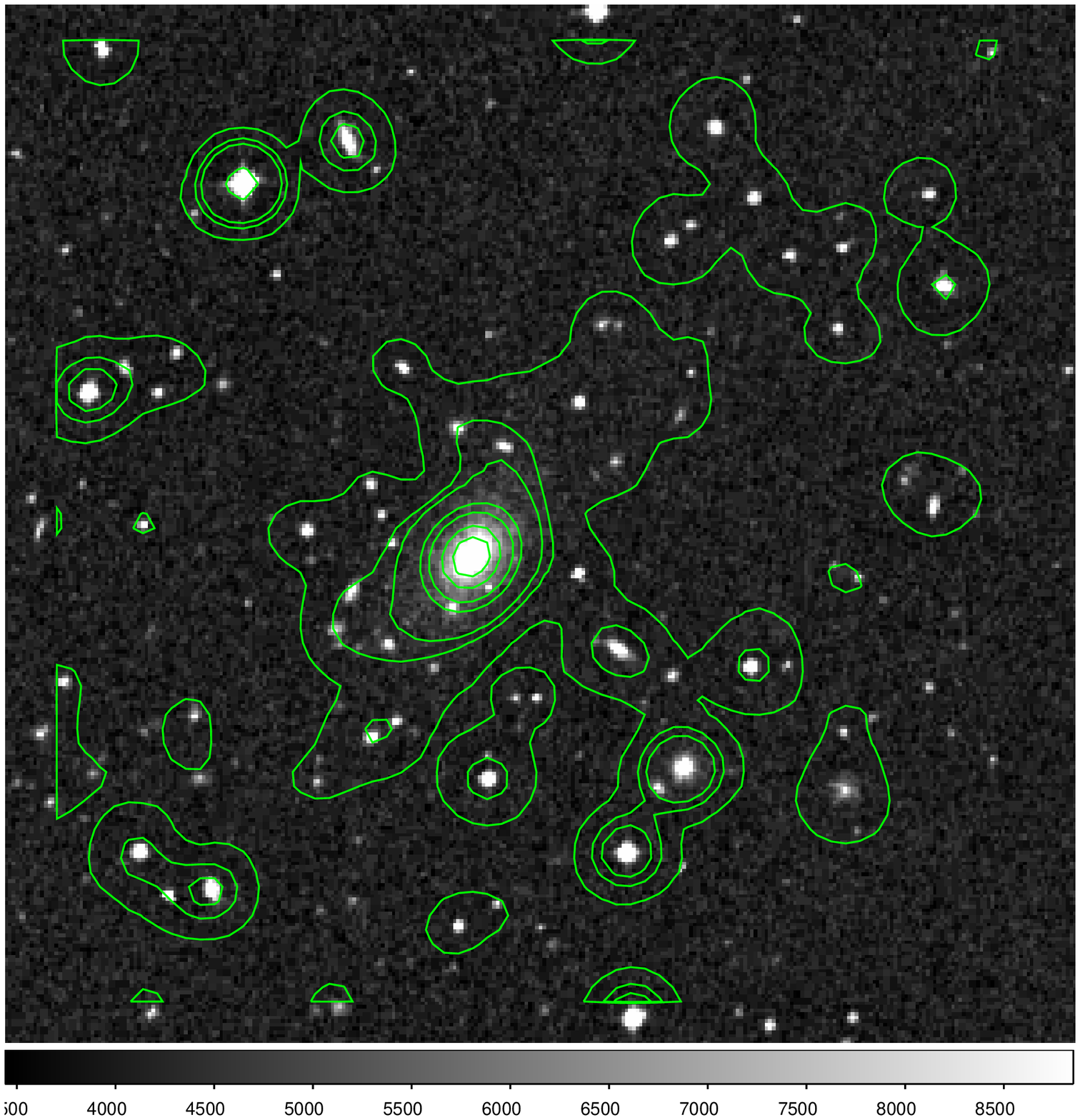}
\figcaption[dss_with_cont.ps]
{Contour map of DSS - R image. An asymmetric galaxy distribution with possible
subcondensations is seen, which is typical of a poor, irregular cluster.\label{fig1}}

\plotone{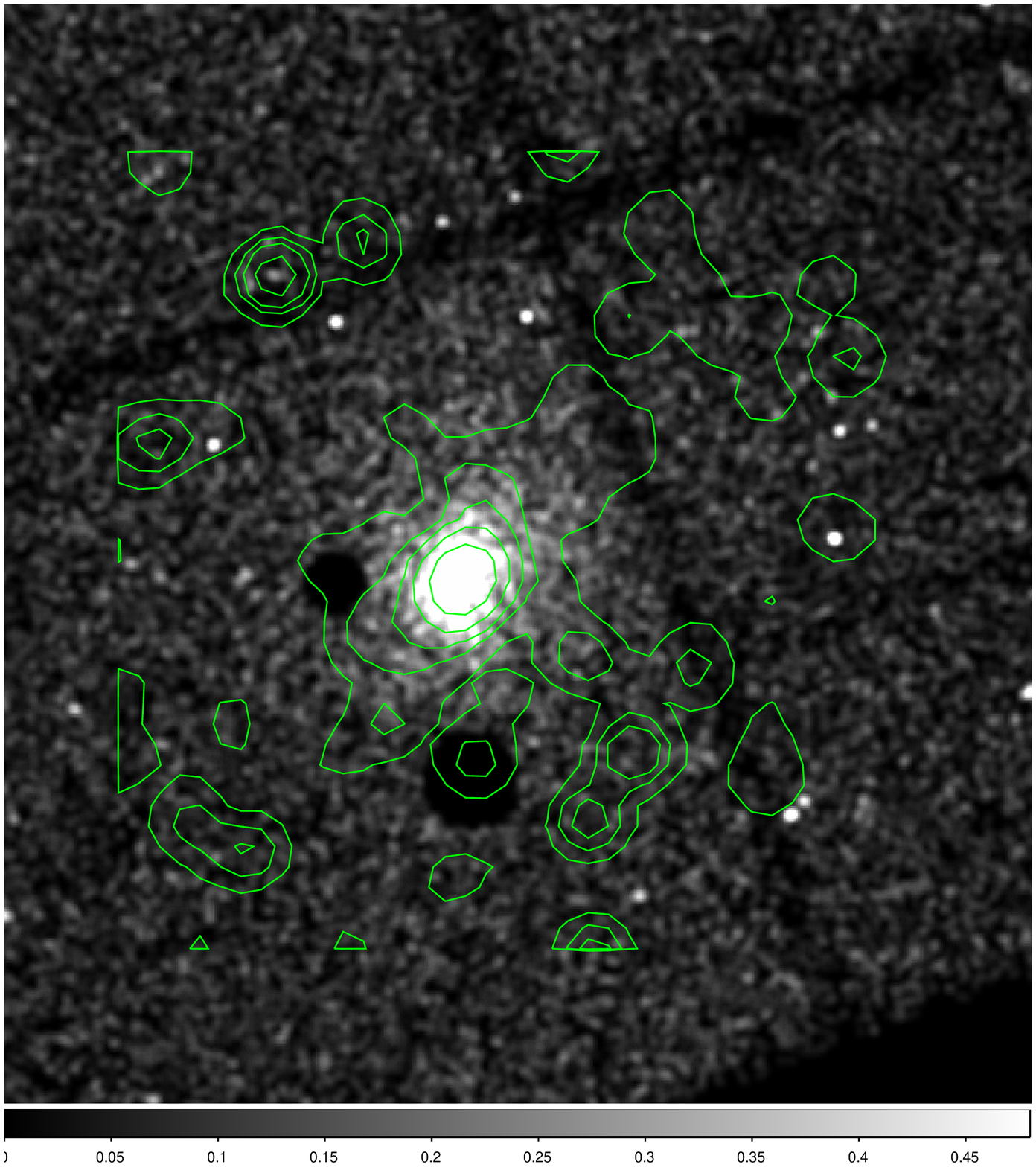}
\figcaption[dss_galdisp_smooth_99.ps]
{The DSS - R is shown in contours overlaying the X-ray image. There are 16 catalogued galaxies 
projected within the X-ray clump.\label{fig2}}

\plotone{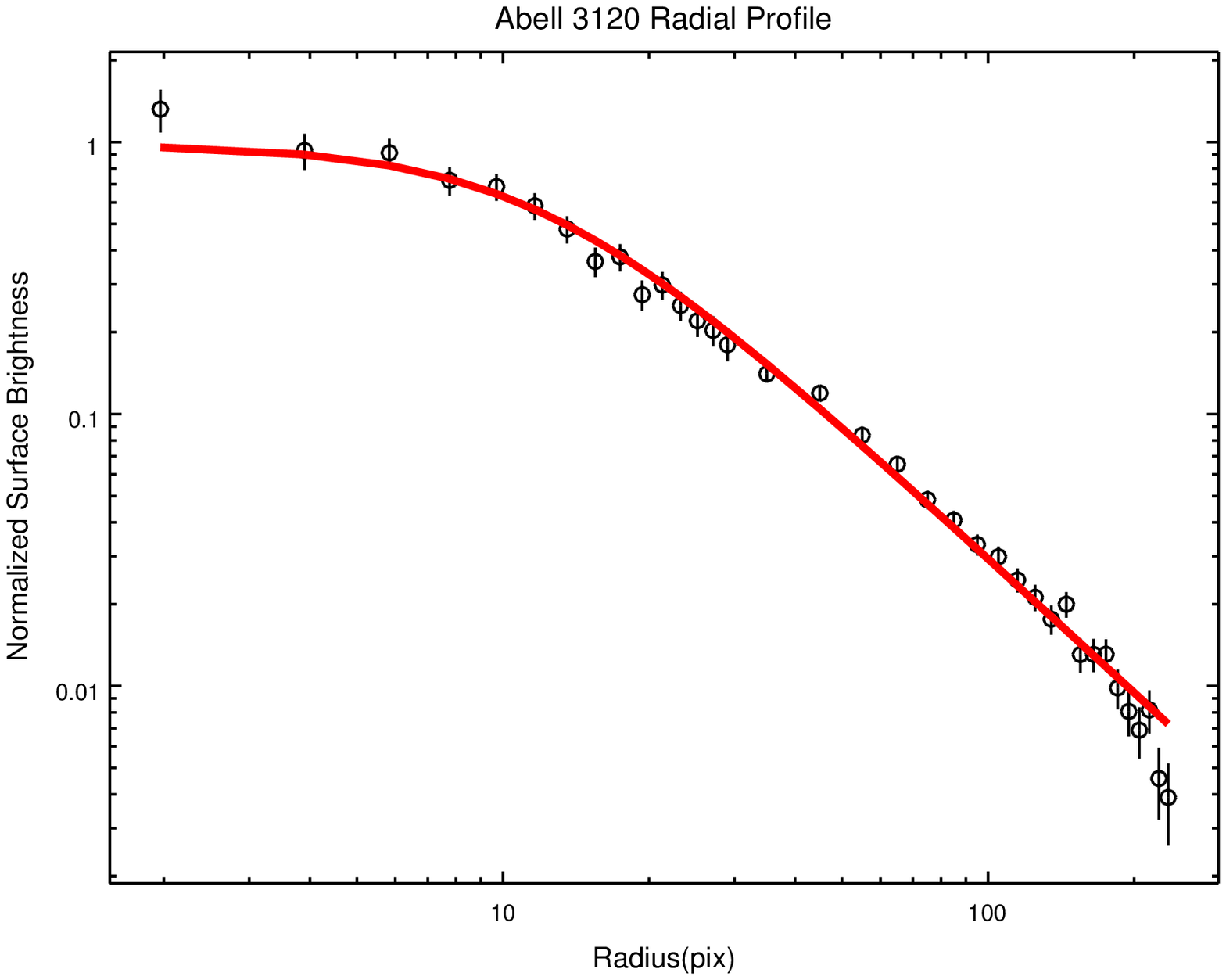}
\figcaption[most_rprofile.ps]
{The radial profile shows the emission is confined to 158 kpc. There may be a slight excess in the center.\label{fig3}}

\plotone{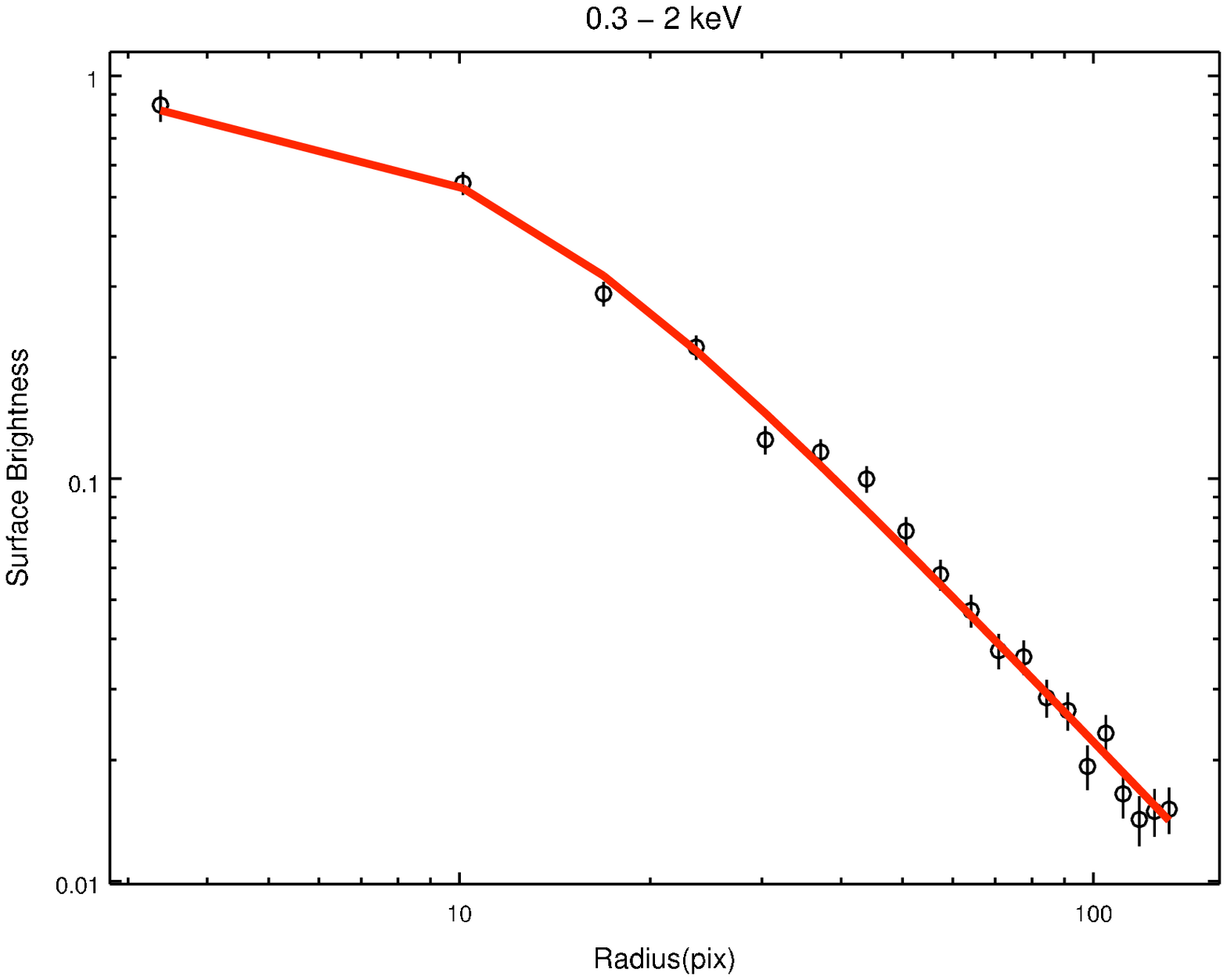}
\figcaption[0.3-2_radial_new.ps]
{The radial profile in the 0.3 - 2 keV band.\label{fig4}}

\plotone{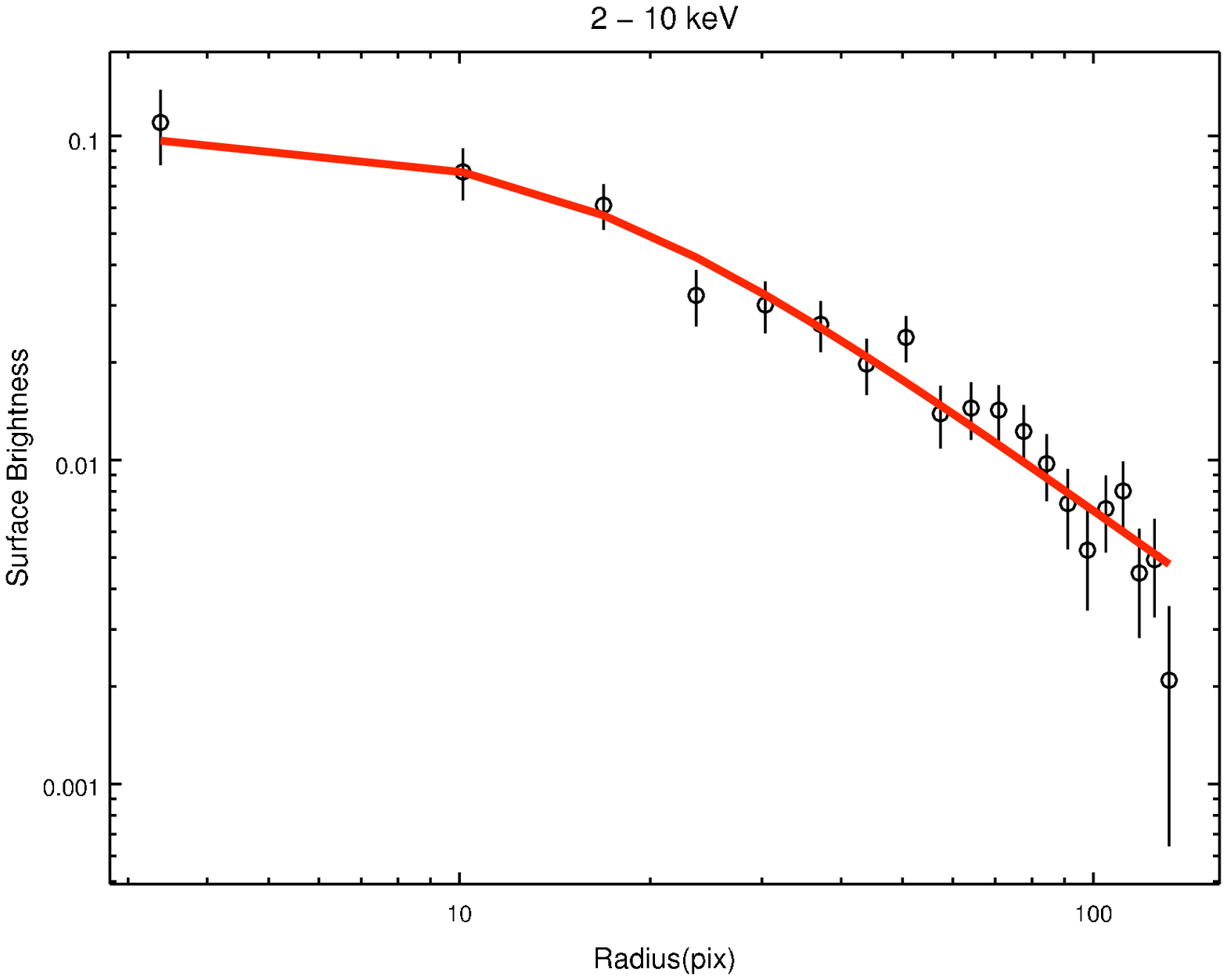}
\figcaption[2-10_radial_new.ps]
{The radial profile in the 2 - 10 keV band appears more extended than the soft band.\label{fig5}}

\plotone{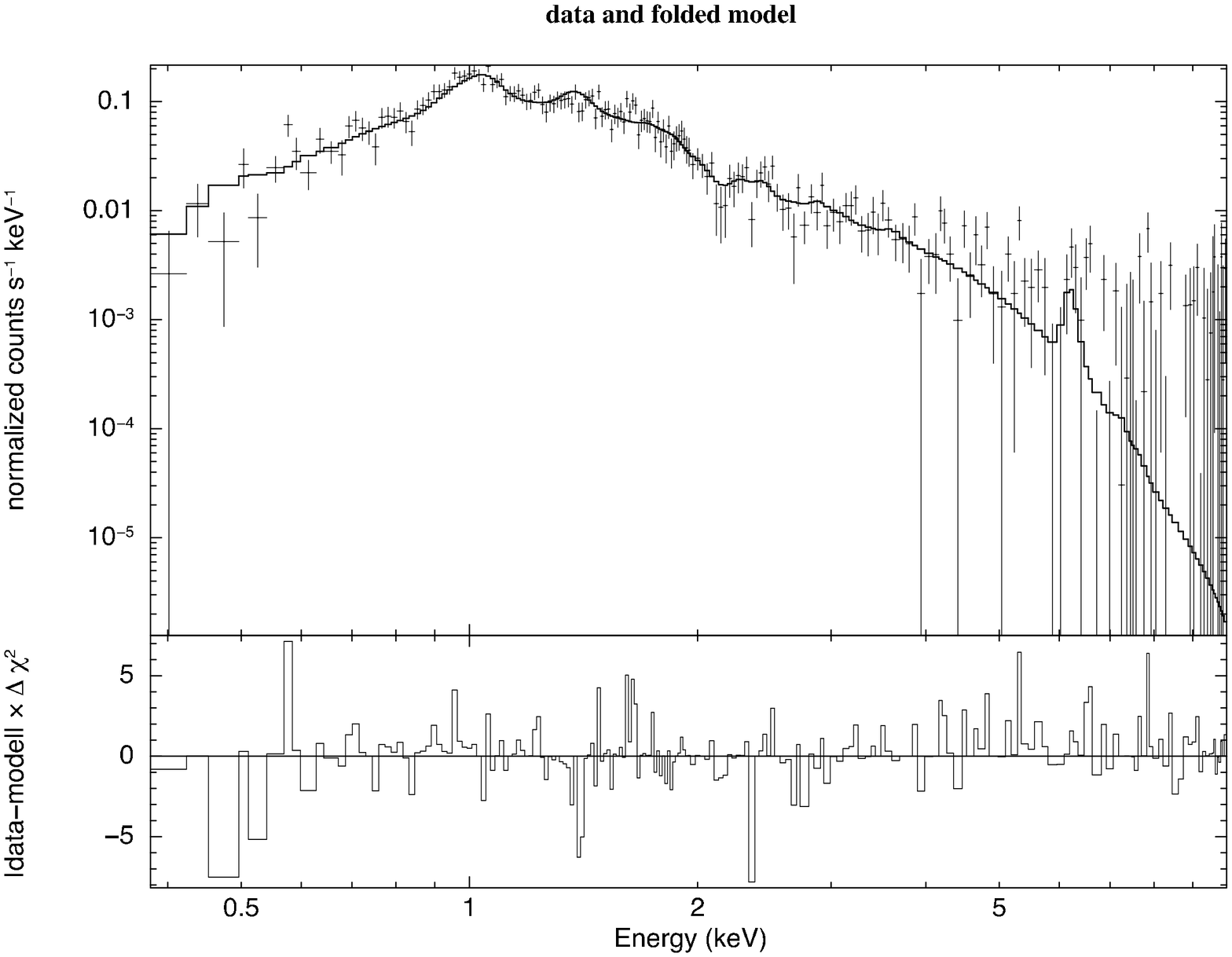}
\figcaption[1rs.ps]
{Data from 0.3 - 10 keV is fit by a single thermal, Raymond \& Smith model. Residuals are
shown below.\label{fig6}}

\plotone{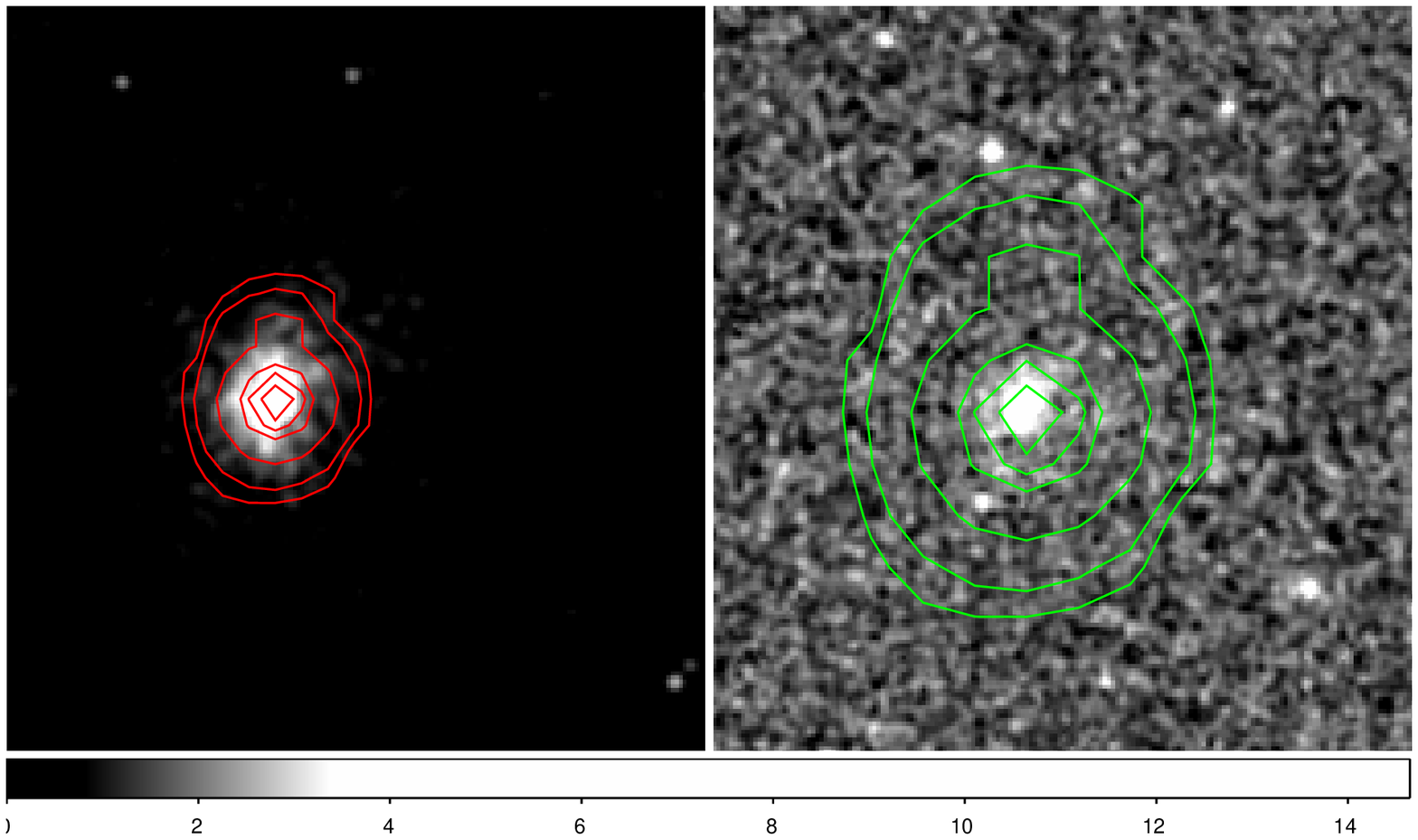}
\figcaption[sumss_con_on_acis_k.ps]
{Radio contours are overlaid on the X-ray
and K band images.\label{fig7}}

\plotone{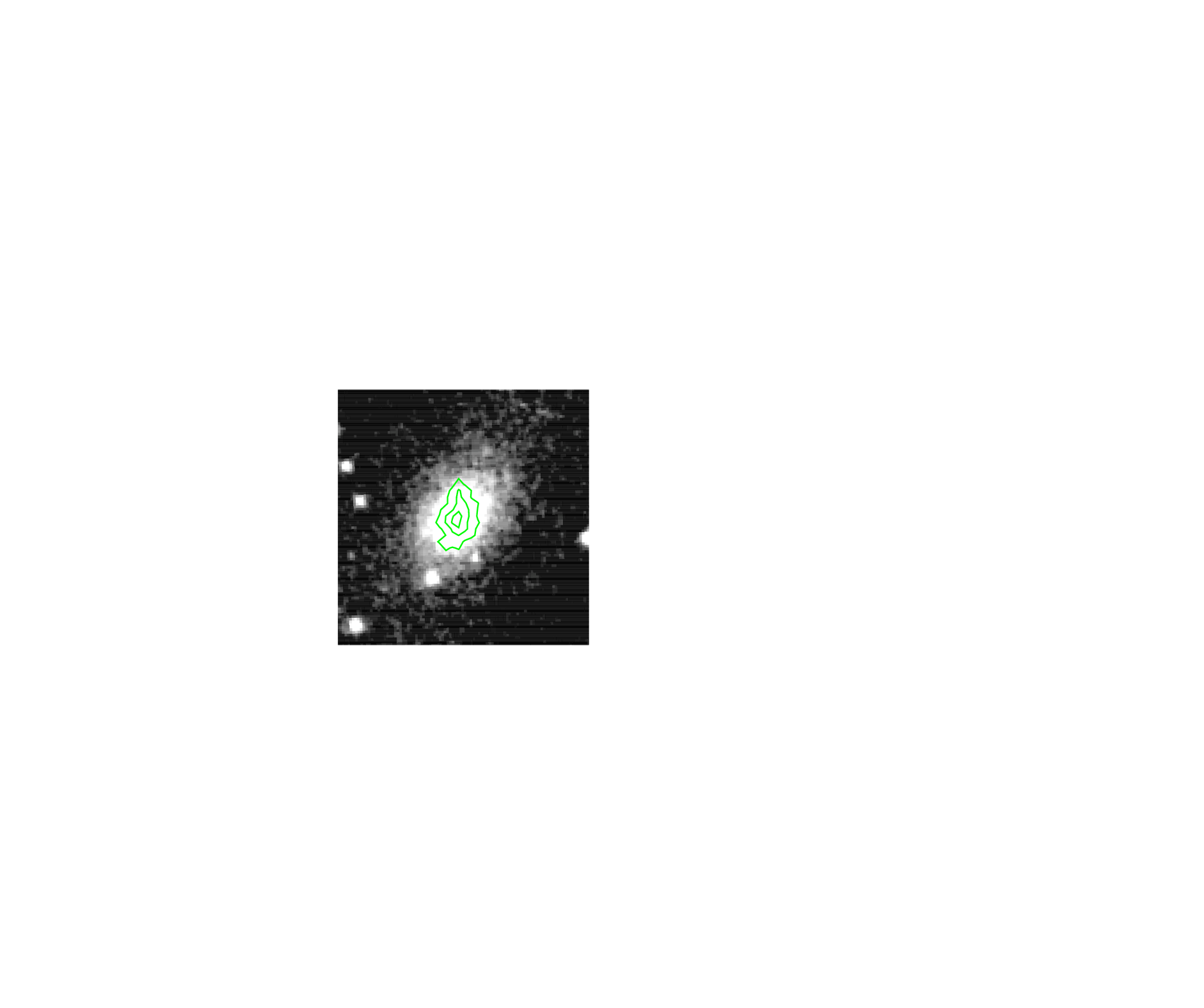}
\figcaption[opt_core.ps]
{X-ray contours are overlayed on the optical image showning the core of the galaxy. A misalignment of the
X-ray contours with the stellar light distribution indicates a possible jet to the North.\label{fig8}}


\end{document}